\documentclass[11pt,a4]{article} 
\usepackage{epsfig}     
\usepackage{amsmath}    
\usepackage{amssymb}   

\newcommand{\beeq}{\begin{eqnarray}}     
\newcommand{\eeeq}{\end{eqnarray}}     
\newcommand{\be}{\begin{equation}}     
\newcommand{\ee}{\end{equation}}

\def\rb{{\bf r}}      
\def\rbp{{\bf r}^\prime}  
\def\bb{{\bf b}}      
\def\lb{{\bf l}}

\def\asb{\overline{\alpha}_s}      
      
\def\gev{\mbox{\rm GeV}}      
\newcommand{\hep}[1]{{\tt hep-ph/#1}}     
\newcommand\appb[3]{{\it Acta Phys. Polon. }{\bf B #1} (#2) #3}

\newcommand\npb[3]{{\it Nucl. Phys. }{\bf B #1} (#2) #3}     
\newcommand\npa[3]{{\it Nucl. Phys. }{\bf A #1} (#2) #3}     
\newcommand\jhep[3]{{\it JHEP }{\bf #1} (#2) #3}     
\newcommand\plb[3]{{\it Phys. Lett. }{\bf B #1} (#2) #3}     
     
\newcommand\epjc[3]{{\it Eur. Phys. J.}{\bf C #1} (#2) #3}     
\newcommand\prd[3]{{\it Phys. Rev. }{\bf D #1} (#2) #3}     
\newcommand\prl[3]{{\it Phys. Rev. Lett. }{\bf  #1} (#2) #3} 
     
\newcommand\zpc[3]{{\it Z. Phys. }{\bf C #1} (#2) #3}     
\newcommand\app[3]{{\it Acta Phys. Polon.}{\bf B #1}(#2) #3}  
      
\begin{document}

\titlepage      
\begin{flushright}     
DESY-04-139\\   
August 2004      
\end{flushright}      
      
\vspace*{1in}      
\begin{center}      
{\Large \bf The unintegrated gluon distribution from the modified BK equation}\\      
\vspace*{0.4in}      
K. \ Kutak$^{(a,c)}$      
and A.\ M. \ Sta\'sto$^{(b,c,d)}$ \\      
\vspace*{0.5cm}       
{\it $^{(a)}$ II.\ Institut f\"ur Theoretische Physik,      
    Universit\"at Hamburg, Germany\footnote{Permanent address}}\\      
$^{(b)}$ {\it Nuclear Theory Group, Brookhaven National Laboratory, Upton, NY 11973, US\footnote{Permanent address}} \\      
$^{(c)}${\it Institute of Nuclear Physics, Radzikowskiego 152,      
 Krak\'ow, Poland}\\  
$^{(d)}$ {\it Theory Group, DESY, Notkestrasse 85, 22607 Hamburg, Germany}
 \vskip 2mm      
\end{center}      
\vspace*{1cm}      
\centerline{(\today)}      
      
\vskip1cm      
\begin{abstract}    
We investigate the recently proposed nonlinear equation for the unintegrated gluon distribution function which includes  the 
subleading effects at small $x$. We obtained numerically the solution to this equation in $(x,k)$ space,  and also the integrated gluon density. The subleading effects affect strongly the normalization and  the  $x$ and $k$ dependence of the  gluon distribution. We show that the saturation scale  $Q_s(x)$, which is obtained from this model, is consistent with the one used in the saturation model by Golec-Biernat and W\"usthoff. We also estimate the nonlinear effects by looking at the relative normalization of the solutions to the  linear and nonlinear equations. It turns out  that the differences are quite large even in the nominally dilute regime, that is when $Q^2 \gg Q_s^2$.
Finally, we calculate the dipole-nucleon cross section.
\end{abstract} 
\newpage
\section{Introduction}  
  
The knowledge of the  QCD dynamics at high energies is essential in understanding 
the hadronic interactions studied at current (HERA, Tevatron) and future 
(LHC) accelerators.  Parton distributions extracted from HERA $ep$ collider 
will be  used in the description of the hadronic processes 
studied at LHC. It is very important to know these parton distributions 
with very high accuracy and, perhaps what  even more important, to be able 
to estimate possible uncertainties which may emerge when extrapolating  
to the kinematic regime  of LHC. A lot of effort is currently devoted 
to extracting the  parton distributions with very high precision \cite{MRST,CTEQ} 
from the available experimental data. 
 
In principle two different frameworks can be used for calculating the  parton distributions. 
 The  standard one is based on the DGLAP evolution and collinear factorization. 
In the high energy limit it is also possible to use  the $k_T$ factorization \cite{CCH} 
in which the QCD interaction is described in terms of the quantity which depends on the transverse momentum of the gluon i.e. the  unintegrated gluon distribution. An  equation  which governs the evolution of  this distribution is the BFKL equation \cite{BFKL}. Its well known solution leads to a very strong  power growth of the gluon density with energy: $\sim s^{\lambda}$ where $\lambda = 4\ln 2 \alpha_s N_c / \pi$ is the BFKL intercept in the leading logarithmic approximation in powers of $\alpha_s \ln 1/x$  (LLX). Next to leading order corrections to BFKL \cite{NLLBFKL} decrease the rate of growth but do not change the power behavior of the gluon distribution.  Thus the growth of the resulting  hadronic cross section 
has to be eventually tamed in order to satisfy the unitarity bound \cite{Froissart}. 
 
The perturbative  parton saturation,  first discussed in a pioneering paper \cite{GLR}, is a phenomenon 
which slows down the rapid growth of the partonic densities.
It is believed that  it leads to the restoration   
of the unitarity of the scattering matrix\footnote{Unitarity or Froissart bound 
is valid with respect to the whole QCD, whereas parton saturation is a perturbative mechanism. As discussed in \cite{KW}, apart from the saturation,  the confinement is also needed to satisfy 
Froissart bound in QCD.}. 
When the density of gluons becomes very high, the gluon recombination 
has also to be taken into account.  This leads to a  modification 
of the evolution equations and their solution results in the saturation of the gluon density.  
In the high energy limit, the parton saturation is described as an infinite 
hierarchy of the coupled evolution equations for the correlators of Wilson lines \cite{Bal}. It is  equivalent  to the JIMWLK 
functional equation \cite{JIMWLK} derived within  the theory of the Color Glass Condensate \cite{CGC}. In the absence of correlations, the first equation in the Balitsky hierarchy decouples, and is then equivalent to the equation derived independently by Kovchegov \cite{Kov} within the dipole formalism \cite{Mueller94}.  
 
It is desirable to have a formalism which  embodies the  resummation 
of the subleading corrections in $\ln 1/x$ and still contains  the
saturation effects. Various attempts in this direction already exist, 
see for example 
\cite{GBMS}-\cite{GLMN}. 
In this contribution  we analyse in better detail the nonlinear equation for the unintegrated gluon distribution function proposed in \cite{KKM01,KK03}.  Its linear term is formulated within the unified BFKL/DGLAP framework \cite{KMS} and the nonlinear term is 
taken from the Balitsky-Kovchegov equation.  
We find that the subleading corrections play an important role in the 
calculation of the unintegrated gluon density since they reduce the value of the intercept and lower the overall normalization of the solution.
We also study the differences of the  solutions in the linear and nonlinear case. 
It is interesting that these differences  become amplified in the case of the integrated gluon density $xg(x,Q^2)$. 
The behaviour of the saturation scale is controlled by the value close to the 
intercept of the solution of the linear equation.  
In our case this  
value is equal to the one suggested by HERA data, $Q_s \sim \exp(\lambda Y)$ with 
 $\lambda \simeq 0.3$.  However, a  more detailed analysis shows that even though 
saturation scale seems to be rather low, the actual numerical differences 
between the linear and nonlinear solutions are much bigger. In other words,  
the effect of nonlinearity on the overall normalization of the solution 
 can be present 
already at the scales exceeding the saturation scale.

The outline of the paper goes as follows: in the next Section we introduce 
the Balitsky-Kovchegov equation and  the formalism which enables to write it in terms of the unintegrated gluon distribution. In Sec.~\ref{sec:f} 
we the  recall basic ingredients of the unified BFKL/DGLAP framework 
 and, following \cite{KKM01,KK03} we formulate the modified Balitsky-Kovchegov equation. 
 
In Sec.~\ref{sec:num} we perform numerical analysis of this equation. We present  
its solution, i.e. the unintegrated gluon distribution as well as the integrated gluon density. Then, we perform the analysis of the saturation scale $Q_s(x)$ and try to quantify 
the importance of the saturation effects by looking at the difference between the linear and nonlinear solution.  
In Sec.~5  we present  the results for the dipole cross section $\sigma(x,r)$ and compare them to the Golec-Biernat and W\"usthoff \cite{GBW} parametrisation. 
 We summarize our study in the last section.

\section{The Balitsky-Kovchegov equation}  
\label{sec:bk} 
In the dipole picture \cite{NikZak} one can view the deep inelastic scattering process as a formation of the $q\bar{q}$  dipole, followed by the  scattering  
of this dipole on the target. In the  high energy,  $s\gg Q^2 \gg \Lambda_{\rm QCD}^2$,  regime  these two processes are factorized, and the total $\gamma^* N$  cross section\footnote{$N$ being a target, nucleon or nucleus.}  can be written as  
\be  
\sigma^{\gamma^* N}_{T,L}(x,Q^2) \; = \; \int \, d^2{\bb} \, d^2{\rb} \, dz \, |\Psi_{T,L}(\rb,Q^2,z)|^2 \, N(\rb,\bb,x)\; ,  
\ee  
where $-Q^2=q^2$ is the photon virtuality  squared, and $x\simeq Q^2/s$ is the usual Bjorken variable.   
The quantity $\Psi_{T,L}(\rb,Q^2,z)$ is  the photon wave function which depends on the virtuality $Q^2$ and the size of the dipole $\rb$, as well as the longitudinal fraction $z$ of the photon momenta carried by the quark. Subscripts $T$ and $L$ denote  the transverse and longitudinal polarisation of the incoming photon, respectively. $N(\rb,\bb,x)$  
is the amplitude for the scattering of the dipole  at impact parameter $\bb$ on the target. It contains all the information about the interaction of the dipole with the target.  
  
The Balitsky-Kovchegov (BK) equation \cite{Bal,Kov} is the non-linear equation  
for the amplitude $N(\rb,\bb,x)$   
\begin{multline}  
\frac{\partial N(\rb,\bb,x)}{\partial \ln 1/x} \;  =\;    
\asb \, \int \, \frac{d^2 \rbp \rb^2}{(\rbp+\rb)^2(\rbp)^2}  
\, \bigg[ \, N(\rbp,\bb+\frac{\rbp+\rb}{2},x)+N(\rbp+\rb,\bb+\frac{\rbp}{2},x)\\  
-N(\rb,\bb,x) -  
 N(\rbp,\bb+\frac{\rbp+\rb}{2},x)\,N(\rbp+\rb,\bb+\frac{\rbp}{2},x)\, \bigg] \;,  
\label{eq:kov}  
\end{multline}  
where $\asb \equiv \frac{\alpha_s N_c}{\pi}$.  
The linear term on the right-hand-side  of (\ref{eq:kov}) is equivalent to the BFKL equation  
in the coordinate space, whereas the nonlinear term is responsible for the  
gluon recombination.   It has been shown \cite{LiVa} to be equivalent to the
triple Pomeron vertex \cite{3Pom}.   
This equation has been independently derived in the dipole picture \cite{Kov} and  
from the Wilson's operator expansion \cite{Bal}. In the latter case,
the equation (\ref{eq:kov}) is just the first member of the infinite hierarchy  
which decouples in the absence of correlations.

In Eq.~(\ref{eq:kov}) there is a nontrivial interplay between the  two sizes: the dipole size $\rb$  and the impact parameter $\bb$. The exact solution, recently studied in \cite{GBS} (and  in \cite{LevNaf} 
with a modified kernel) has a very complicated $\bb$ and $\rb$ dependence which comes as a consequence of the conformal symmetry of this equation.  
Solutions to this equation  simplified ignoring the impact parameter dependence have been extensively studied both analytically \cite{LevTuch,MuPe,BFL04} and numerically \cite{GBMS,LevLub,Braun,RuWe04,IkMcL04}. 
Here we are interested in the unintegrated gluon density $f(x,k^2)$ averaged over the impact parameter $\bb$ . Following \cite{KKM01,KK03} we  make an ansatz that this dependence  factorizes
\be  
N(\rb,\bb,x) \; = \; n(r,x )\, S(b) \; , 
\label{eq:nfact}  
\ee  
with the normalization conditions on a profile $S(b)$  
\beeq  
\int d^2 {\bb} \, S(b) & =  & 1 \; ,\nonumber \\  
\int d^2 {\bb} \, S^2(b) & =  & \frac{1}{\pi R^2}  \; , 
\eeeq  
where $R$ is the target size in the impact parameter.  
  
We are fully aware that the assumption (\ref{eq:nfact}) is  crude,   
since it implies an approximation of an  infinite and uniform target. To obtain the full $\bb$ dependence one should  
consider the exact equation (\ref{eq:kov}).  
  
One can now transform the equation (\ref{eq:kov}) in the momentum space,
\be  
\Phi(\lb,\bb,x)  =  \int \frac{d^2{\rb}}{2\pi r^2}\, e^{i\lb\rb}\, N(\rb,\bb,x) \; , 
\label{eq:phi}  
\ee  
and taking
\be  
\Phi(\lb,\bb,x)  = \phi(l,x) S(b) \; . 
\label{eq:pfact}  
\ee  
We neglect the angular dependence and assume that the functions $n$ and $\phi$ depend  
only on the absolute values of $r \equiv |\rb|$ and $l \equiv |\lb|$.  
The relations (\ref{eq:nfact},\ref{eq:phi},\ref{eq:pfact}) enable to write the equation (\ref{eq:kov}) in the   
following form \cite{Kov} 
\be  
\frac{\partial{\; \phi(l,x)}}{\partial{\; \ln 1/x}} \; = \asb   
\bigg[ K \otimes  \phi - \frac{1}{\pi R^2}  \phi^2(l,x)\bigg] \; , 
\label{eq:JKKK} 
\ee  
where we have integrated both sides of (\ref{eq:kov}) over $d^2{\bb}$.  
Note that  while $N(\rb,\bb,x)$ and $\Phi(\lb,\bb,x)$ are dimensionless,  
 the  functions $n(r,x)$ and $\phi(l,x)$ have dimension $[\frac{1}{\rm energy^2}]$ due to the definitions (\ref{eq:nfact}) and (\ref{eq:pfact}).  
The operator $K$ is the BFKL kernel \cite{BFKL} in  momentum space in the LLx  
approximation.\\  
Let us now explicitly show how  to find the  relation between $\phi(l,x)$ and the unintegrated gluon distribution $f(x,k^2)$ defined through 
\be 
xg(x,Q^2) \; \equiv \; \int^{Q^2} \frac{dk^2}{k^2} \, f(x,k^2) \; ,
\label{eq:xg} 
\ee 
with $xg(x,Q^2)$ being the  integrated gluon density. 
The unintegrated gluon distribution is related to the dipole cross section   
\be  
\sigma(r,x)={8 \pi^2 \over N_c}\int{dk\over k^3}[1-J_0(kr)]\alpha_s f(x,k^2) \; , 
\label{eq:dipxsec}  
\ee  
which in turn can be obtained from the amplitude $N(\rb,\bb,x)$ by performing the integration over $\bb$ 
\be  
\sigma(r,x)=2\int d^2{\bf b} \, N({\bf r},{\bf b},x)  \, . 
\label{eq:dipxsecn} 
\ee  
Using (\ref{eq:phi}),(\ref{eq:dipxsec}) and (\ref{eq:dipxsecn})   
one obtains  
\be  
\phi(l,x)=\frac{1}{2}\int \frac{d^2{\rb}}{2\pi r^2}\,  
e^{i\lb\rb}\,\frac{8\pi^2}{N_c}\int\frac{dk}{k^3}[1-J_{0}(kr)]\alpha_{s}f(x,k^2)  \; . 
\ee  
Integrating over angles yields  
\be  
\phi(l,x)=\frac{2\pi^2}{N_c}\int_{l^2}^{\infty}\frac{dk^2}{k^4}\int_{0}^{\infty}\frac{dr}{r}J_{0}(lr)[1-J_{0}(kr)]  
\alpha_{s}f(x,k^2) \; , 
\ee  
and the integral over $r$ gives  
\be  
\phi(l,x)={\pi^2 \over N_c}\int_{l^2}^{\infty}{dk^2\over  
k^4}\ln\left(\frac{k^2}{l^2}\right) \alpha_s f(x,k^2)  \; . 
\ee  
Now we need  to invert the  operator 
\be  
\hat{O} = \frac{\pi^2 \alpha_s} {N_c}\int_{l^2}^{\infty}{dk^2\over  
k^4}\ln\left(\frac{k^2}{l^2}\right) g(k^2) \; , 
\ee  
(where $g(k^2)$ is a test function) 
to get the expression for $f(x,k)$. 
Multiplying both sides of  (12) by $l^2$ and performing the Mellin transform  
with respect to $l^2$ we obtain

$$ 
\phi(\gamma,x)\equiv\int dl^2 l^2\phi(l,x)(l^2)^{\gamma-1}=\int dl^2 l^2{\pi^2 \over N_c}\int_{l^2}^{\infty}{dk^2\over  
k^4}\ln\left(\frac{k^2}{l^2}\right) \alpha_s f(x,k^2)(l^2)^{\gamma-1}=
$$
\be 
=\frac{\alpha_s\pi^2}{N_c}f(\gamma)\frac{1}{(\gamma+1)^2}, 
\ee  
and equivalently  
\be  
f(\gamma)=\frac{N_c}{\alpha_s\pi^2}(\gamma+1)^2\phi(\gamma,x) \; . 
\ee  
The inverse Mellin transform gives  
\be  
f(x,l^2)=\frac{N_c}{\alpha_s\pi^2}\int\frac{d\gamma}{2\pi  
i}(l^2)^{-\gamma}(1+\gamma)^2\phi(\gamma,x)=\frac{N_c}{\alpha_s\pi^2}(1-l^2\frac{d}{dl^2})^2l^2\phi(l,x)\; .   
\label{eq:relacja} 
\ee  
This relation between functions  $f$ and $\phi$ has been first derived  in \cite{Braun} and  also on in \cite{KKM01,KK03}. 
\section{The non-linear equation for the unintegrated density}  
\label{sec:f} 
The relation  (\ref{eq:relacja}) allows us to transform (\ref{eq:JKKK}) 
into an equation  for the unintegrated gluon distribution  
  
$$  
\frac{\partial f(x,k^2)}{\partial \ln 1/x}=  \frac{\alpha_s N_c}{\pi}
\,k^2\int_{k_0^2}{dk^{\prime 2}\over  
k^{\prime 2}}\left \{{f\left({x},k^{\prime 2}\right)-f\left({x},k^2\right)\over |k^{\prime 2}-k^2|} +  
{f\left({x},k^2\right)\over  
[4k^{\prime 4}+k^4]^{{1\over 2}}}\right\}  
$$  
\begin{equation}  
-\alpha_s\left(1-k^2{d\over dk^2}\right)^2{k^2\over R^2}  
\left[\int_{k^2}^{\infty}  
{dk^{\prime 2}\over k^{\prime 4}}\ln\left(  
{k^{\prime 2}\over k^2}\right)f(x,k^{\prime 2})\right]^2 \; . 
\label{eq:modf}  
\end{equation}  
It is BFKL equation supplemented by the negative nonlinear term.  
\subsection{A partial resummation of the NLLx corrections}  

Equation (\ref{eq:modf}) contains the BFKL kernel at leading logarithmic (LLx) accuracy.   
This is a coarse approximation as far as a description of the HERA data is concerned. It is well known \cite{NLLBFKL} that the  NLLx corrections 
to the BFKL equation are quite large. To make the equation  more realistic,  
it was proposed  \cite{KKM01,KK03} to   
implement in the linear term  of (\ref{eq:modf}) a unified BFKL-DGLAP framework developed in  \cite{KMS}.   
In this scheme \cite{KMS}, the BFKL kernel becomes modified by  the  
consistency constraint \cite{AGKS,KMScc}  
\be  
k'^2 \; < \; k^2 / z \; , 
\label{eq:kincon}  
\ee  
imposed  onto the real-emission part of the kernel in Eq.~(\ref{eq:modf}) 
\be 
\int \frac{dk'^2}{k'^2} \, 
\bigg\{ \, \frac{f(\frac{x}{z},k^{\prime 2}) \, \Theta(\frac{k^2}{z}-k'^2)\, 
-\, f(\frac{x}{z},k^2)}{|k'^2-k^2|}   +\, \frac{f(\frac{x}{z},k^2)}{|4k^{\prime 
4}+k^4|^{\frac{1}{2}}} \, \bigg\} \, . 
\ee 
The consistency constraint (\ref{eq:kincon}) resums a large part of the subleading corrections  coming from a choice of  scales in the BFKL kernel \cite{Salam98,CCS99}. Additionally, the non-singular (in $x$) part of the leading order (LO) DGLAP   splitting function is included into the evolution  
\be  
\int_x^1 \frac{dz}{z} K\otimes f \rightarrow \int_x^1 \frac{dz}{z} K \otimes f + \int^{k^2}\frac{dk'^2}{k'^2} \int_x^1 dz\bar{P}_{gg}(z) f(\frac{x}{z},k^{\prime 2}) \; ,  
\ee  
where  
\be  
\bar{P}_{gg}(z) = P_{gg}(z) - \frac{2N_c}{z} \; .  
\ee  
Additionally, we assume that in our evolution equation $\alpha_s$  runs with scale $k^2$ which is  yet another source of important NLLx corrections. 
The final improved nonlinear equation for the unintegrated gluon density is as follows  
\begin{multline}   
f(x,k^2) \; = \; \tilde f^{(0)}(x,k^2) + \\   
+\,\frac{\alpha_s(k^2) N_c}{\pi} k^2 \int_x^1 \frac{dz}{z} \int_{k_0^2}
\frac{dk'^2}{k'^2} \,   \bigg\{ \, \frac{f(\frac{x}{z},k^{\prime 2}) \,
\Theta(\frac{k^2}{z}-k'^2)\,   -\, f(\frac{x}{z},k^2)}{|k'^2-k^2|}   +\,
\frac{f(\frac{x}{z},k^2)}{|4k^{\prime   4}+k^4|^{\frac{1}{2}}} \, \bigg\}\,+
\\  + \, \frac{\alpha_s (k^2) N_c}{\pi} \int_x^1 dz \,   \bar{P}_{gg}(z)
\int^{k^2}_{k_0^2} \frac{d k^{\prime 2}}{k^{\prime 2}} \,  
f(\frac{x}{z},k'^2)\,-  \\  -\left(1-k^2{d\over dk^2}\right)^2{k^2\over
R^2}\int_x^1\frac{dz}{z}\left[\int_{k^2}^{\infty}  {dk^{\prime 2}\over
k^{\prime 4}}\alpha_s(k^{\prime   2})\ln\left(  {k^{\prime 2}\over
k^2}\right)f(z,k^{\prime 2})\right]^2 \; .   \label{eq:fkovres}    
\end{multline}    In \cite{KMS} the inhomogeneous term was defined in terms of
the integrated gluon 
 distribution
\begin{equation}  
\tilde f^{(0)}(x,k^2)=\frac{\alpha_S(k^2)}{2\pi}\int_x^1 
dzP_{gg}(z)\frac{x}{z}g\left(\frac{x}{z},k_0^2\right) 
\label{eq:input} 
\end{equation} 
 taken at scale $k_0^2=1 \gev^2$. This scale was also used as a cutoff in the 
linear version of the evolution equation (\ref{eq:fkovres}). In the linear case this provided a very good description of $F_2$ data with a minimal number 
of physically motivated parameters, see  \cite{KMS}. 
The initial integrated density at scale $k_0^2$ was parametrised as 
\be 
xg(x,k_0^2)=N(1-x)^{\rho} \; ,
\label{eq:gluony} 
\ee 
where $N=1.57$ and $\rho=2.5$. 

Let us finally note that in this model only the linear part of the BK equation has subleading corrections.
We  do not know yet how to include these corrections in the nonlinear term. This would require
the exact knowledge of the triple Pomeron vertex \cite{3Pom} at NLLx accuracy, which is  yet unknown beyond the LLx approximation.

\section{Numerical analysis}  
 
\label{sec:num} 
\subsection{The unintegrated and integrated gluon density}  

In this section we recall the method of solving Eq.~(\ref{eq:fkovres}) and we 
present the  numerical results for the unintegrated gluon distribution function  
$f(x,k^2)$ and the integrated  gluon density $xg(x,Q^2)$. The method of solving 
(\ref{eq:fkovres}), developed in \cite{KK03},  relies on reducing it to an 
effective evolution equation in $\ln 1/x$
 with the boundary condition 
at some moderately small value  of $x$ (i.e. $x=x_0 \sim 0.01$).   
 
To be specific,  we make the following approximations:  
\begin{enumerate} 
\item The  consistency constraint $\Theta(k^2/z-k^{\prime 2})$ in the  BFKL kernel  is replaced by the following  
effective ($z$ independent)  term  
\begin{equation} 
\Theta(k^2/z-k^{\prime 2}) \rightarrow \Theta(k^2-k^{\prime 2}) + 
\left({k^2\over k^{\prime 2}}\right)^{\omega_{eff}}\Theta(k^{\prime 2}-k^2)\,.
\label{bfkleff} 
\end{equation} 
This  is motivated by the structure of the consistency constraint  
in the moment space, i.e.  
\begin{equation} 
\omega \int_0^1{dz \over z}z^{\omega}\Theta(k^2/z-k^{\prime 2})=\Theta(k^2-k^{\prime 2}) + 
\left({k^2\over k^{\prime 2}}\right)^{\omega}\Theta(k^{\prime 2}-k^2) \; , 
\label{ccmom} 
\end{equation} 
\item The splitting function is approximated in the following way 
\begin{equation} 
\int_x^1{dz\over z}[zP_{gg}(z)-2N_c] 
f\left({x\over z},k^{\prime 2}\right) \rightarrow \bar P_{gg}(\omega=0) f(x,k^{\prime 2}) \; , 
\label{dglapeff} 
\end{equation} 
where $\bar P_{gg}(\omega)$ is a moment function  
\begin{equation} 
\bar P_{gg}(\omega)=\int_0^1{dz\over z}z^{\omega}[zP_{gg}(z)-2N_c] \; , 
\label{pggmom} 
\end{equation} 
and 
\begin{equation} 
\bar{P}_{gg}(\omega=0)=-\frac{11}{12} \; . 
\label{eleventwelve}
\end{equation} 
This approximation corresponds to retaining only  the leading term in the expansion  
of $\bar P_{gg}(\omega)$ around $\omega=0$, see \cite{LOWX}.   
\end{enumerate}

Using these approximations in (\ref{eq:fkovres}) we obtain 
\begin{multline} 
\frac{\partial f(x,k^2)}{\partial \ln (1/x)} \; = \;  \\   
\,\frac{\alpha_s(k^2)N_c}{\pi} \int_{k_0^2} \frac{dk'^2}{k'^2} \,  
\bigg\{ \, \frac{f\left({x},k^{\prime 2}\right)\left [\Theta\left(k^2-k^{\prime 2} 
\right) +\left(\frac{k^2}{k^{\prime 2}}\right)^{\omega_{\rm eff}}\Theta\left(k^2-k^{\prime 2}\right)\right] - f\left(x,k^2\right)}{|k'^2-k^2|}    
+\\+\, \frac{f(x,k^2)}{|4k^{\prime  
4}+k^4|^{\frac{1}{2}}} \, \bigg\}\,+ \, \frac{\alpha_s(k^2)N_c}{\pi}  \,  
\bar{P}_{gg}(0) \int^{k^2}_{k_0^2} \frac{d k^{\prime 2}}{k^{\prime 2}} \,  
f(x,k'^2)\,-  \\-\left(1-k^2{d\over dk^2}\right)^2{k^2\over R^2}   
\left[\int_{k^2}^{\infty}  {dk^{\prime 2}\over k^{\prime 4}}\alpha_s(k^{\prime  
2})\ln\left(  {k^{\prime 2}\over k^2}\right)f(x,k^{\prime 2})\right]^2 \; .  
\label{eq:kovev}   
\end{multline}

\begin{figure}[htb]  
\centering{   
\includegraphics[width=0.489\textwidth]{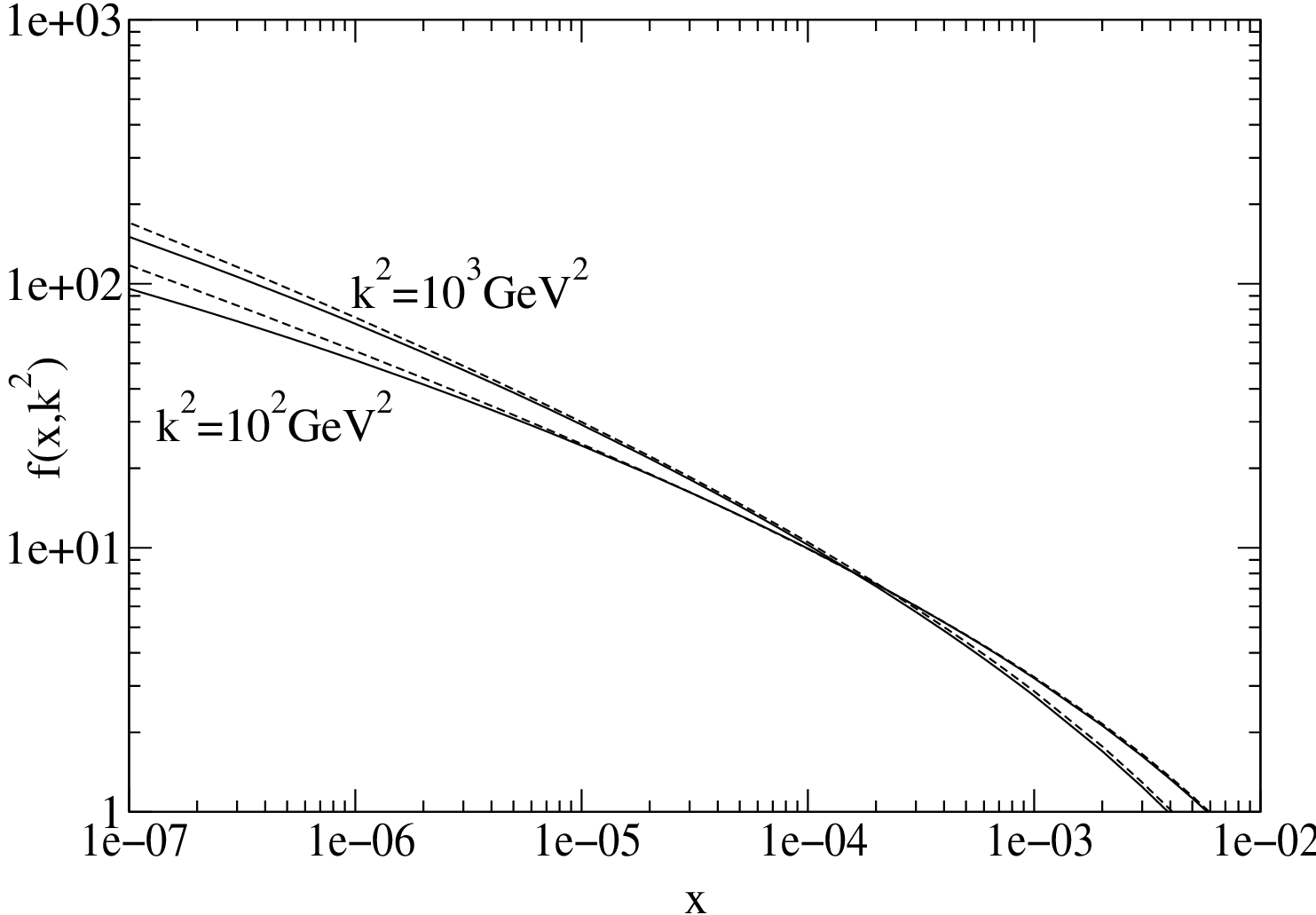}\hfill\includegraphics[width=0.489\textwidth]{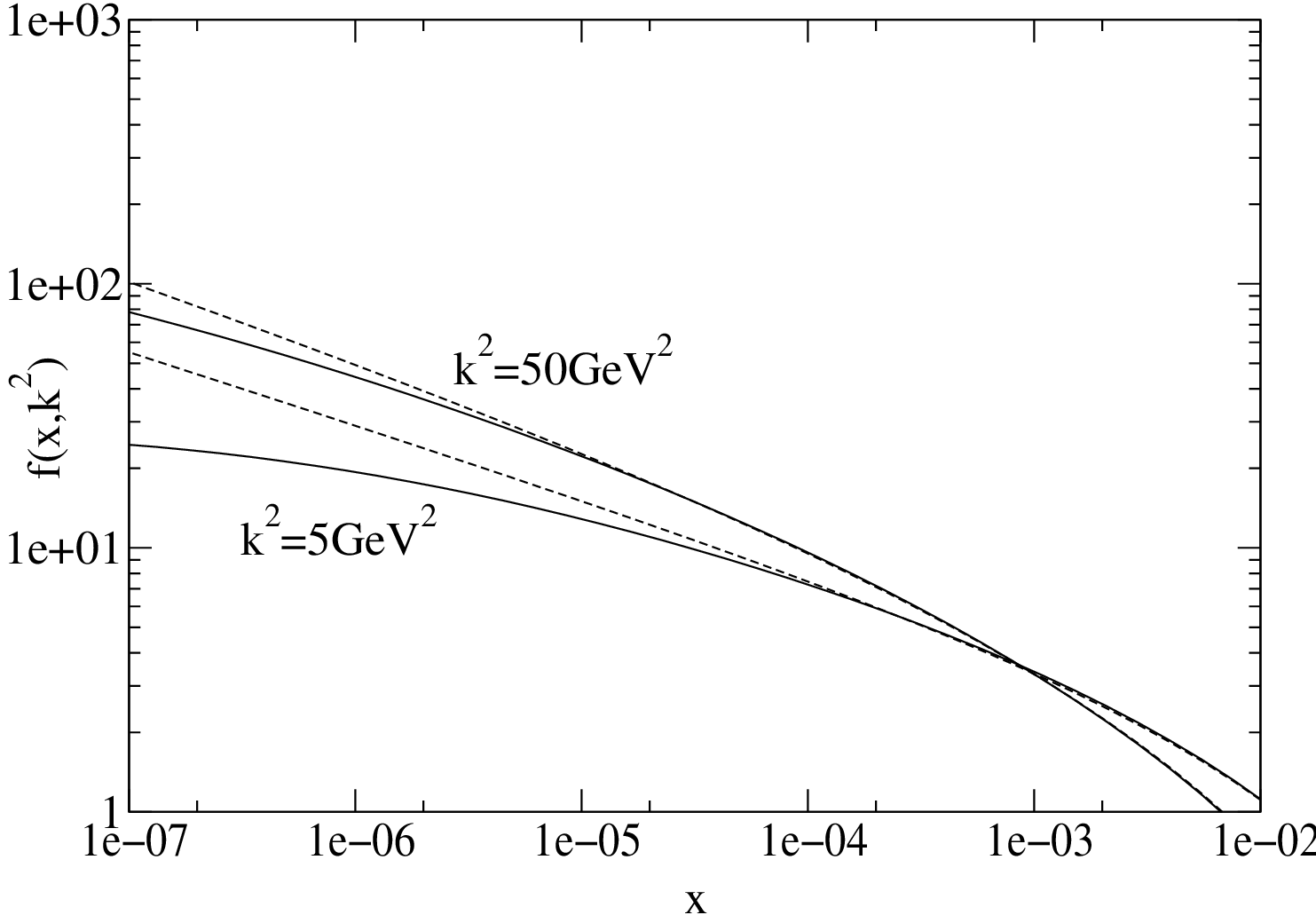}\\}  
\caption{\em The unintegrated 
gluon distribution $f(x,k^2)$ obtained from Eq.(\ref{eq:kovev}) as a  function
of $x$ for  different values $k^2 = 10^2 \ \gev^2$ and $k^2 = 10^3  \ \gev^2$
(left) and for  $k^2 = 5 \ \gev^2$ and $k^2 = 50  \  \gev^2$ (right)  .
The solid lines correspond to the solution of the nonlinear equation
(\ref{eq:kovev}) whereas the dashed lines correspond to the linear  BFKL/DGLAP term
in (\ref{eq:kovev}).} 
\label{fig:fx}
\end{figure} 

 First, the equation (\ref{eq:kovev}) was solved   with the non-linear term
neglected starting from the initial conditions   at $x=10^{-2}$ given by
(\ref{eq:input}).   The parameter $\omega_{\rm eff}$ was adjusted in such a
way   that the solution of the linear part of (\ref{eq:kovev})  matched the
solution of   the original equation in the BFKL/DGLAP framework \cite{KMS}.  
This procedure gives $\omega_{\rm eff}=0.2$      and the solution of the
linear part of (\ref{eq:kovev}) reproduces the   original results of  \cite{KMS}
within     3\% accuracy in the region $10^{-2}>x>10^{-8}$ and $2\
\gev^2<k^2<10^6 \ \gev^2$. This matching procedure has also the advantage that
the quark contribution  present in the original BFKL/DGLAP
framework  is effectively  included by fitting the  value of $\omega_{\rm
eff}$.  The full non-linear equation (\ref{eq:kovev}) was then solved using
the same initial conditions and setting   $R=4 \ \gev^{-1}$.

 In 
Fig.~\ref{fig:fx} we plot the unintegrated gluon distribution function   as a
function of $x$ for different values of $k^2$.   This figure compares  the results of two
calculations, based on the linear  and   nonlinear equations.
 The   differences are not large, however  there is
some suppression due to the   nonlinearity at smallest values of $x 
\le 10^{-5}$. 

\begin{figure}[htb]
\centering{
\includegraphics[width=0.489\textwidth]{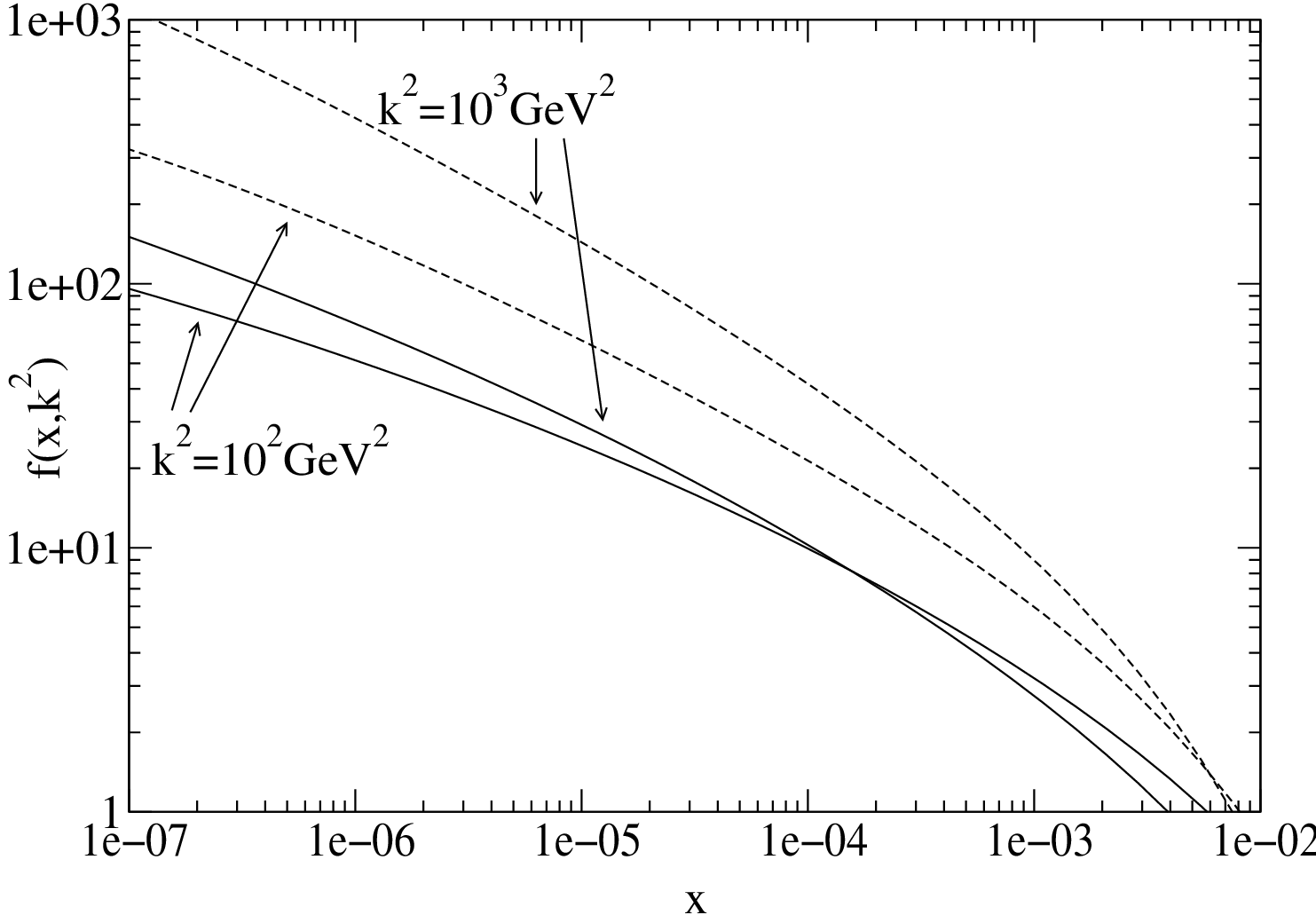}
\hfill\includegraphics[width=0.489\textwidth]{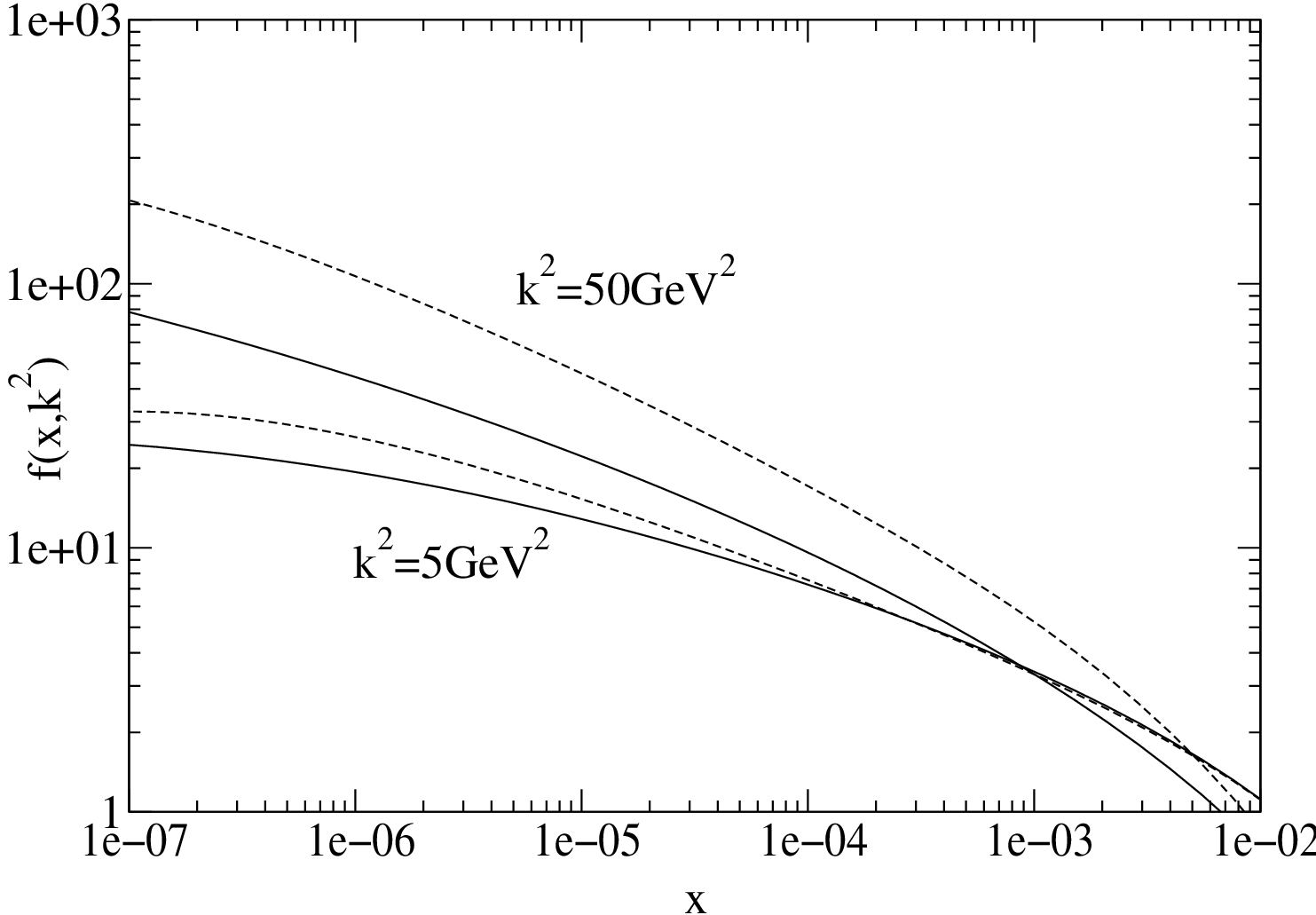}\\}
\caption{The same as Fig.~\ref{fig:fx} but now the modifed BK equation (\ref{eq:kovev}) (solid lines) is compared with the
original BK equation (\ref{eq:modf}) without subleading corrections (dashed lines).}
\label{fig:fxbk}
\end{figure}

The subleading   corrections strongly decrease the value of the
intercept with respect to the   LLx  value and   the
nonlinear term becomes important only at   very low values of $x$. 
As is evident from Fig.~\ref{fig:fxbk} the subleading corrections cause a large suppression
in the normalisation, also at moderate values of $x$. This is due to the fact that
the non-singular in $x$ part   of the $P_{gg}$ splitting function was included
into the evolution. This term is negative and is important at large and moderate values
of $x$.

\begin{figure}[t]  \centering{   
\includegraphics[width=0.49\textwidth]{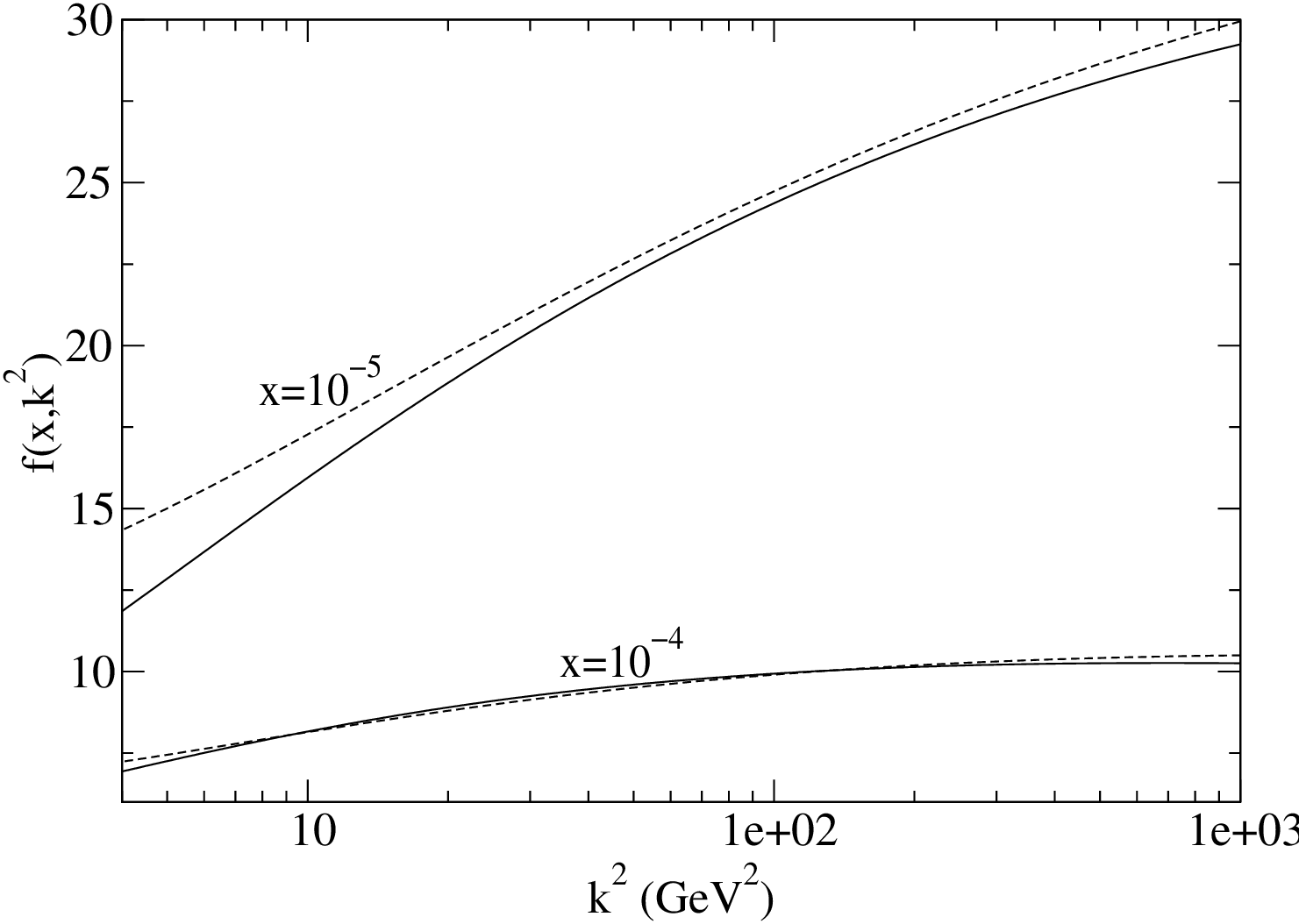}\hfill
\includegraphics[width=0.49\textwidth]{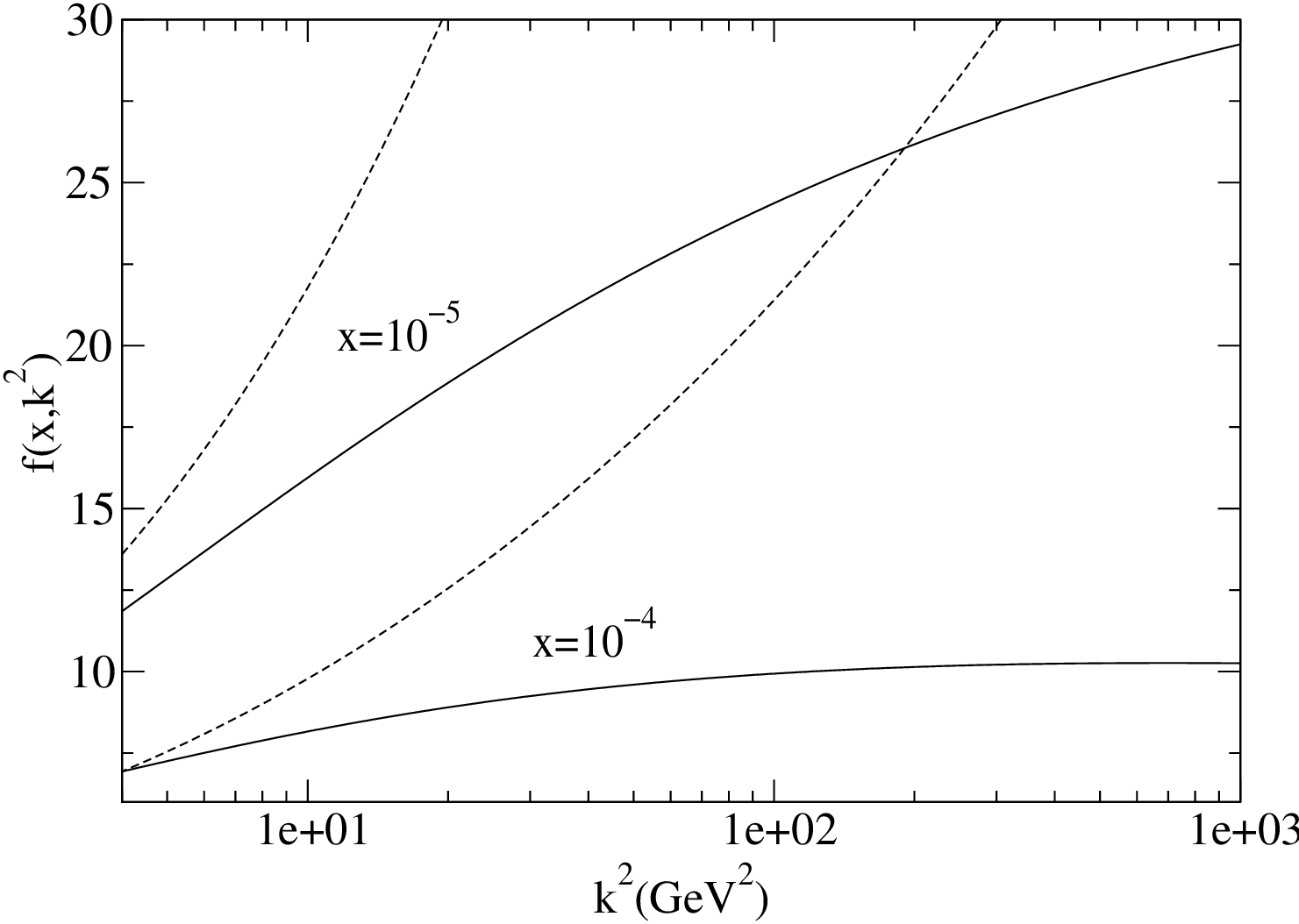}
}
\caption{\em The unintegrated gluon distribution $f(x,k^2)$ 
 as a function of $k^2$ for two values of $x=10^{-5}$
and $10^{-4}$  . 
Left: solid lines correspond to the solution of the nonlinear
equation (\ref{eq:kovev}) whereas dashed lines correspond to linear 
BFKL/DGLAP term in (\ref{eq:kovev}).
Right: solid lines correspond to the solution of the nonlinear equation (\ref{eq:kovev})
whereas dashed lines correspond to the solution of the original BK equation without
the NLLx modifications in the linear part (\ref{eq:modf}).
} 
\label{fig:fk} 
\end{figure}
The same conclusions can be reached by  investigating the plots in Fig.~\ref{fig:fk}
where the unintegrated density is shown as a function of the transverse momentum $k^2$
for fixed values of $x$.
The nonlinear effects seem to have a moderate impact in that region.
On the other hand 
 the subleading corrections are substantial. For example, at $x=10^{-5}$ and $k^2=10 \, {\rm GeV^2}$
 the reduction in magnitude of the unintegrated  gluon density is about  $25\%$.

\begin{figure}[t] 
\centering{ 
 \includegraphics[width=0.489\textwidth]{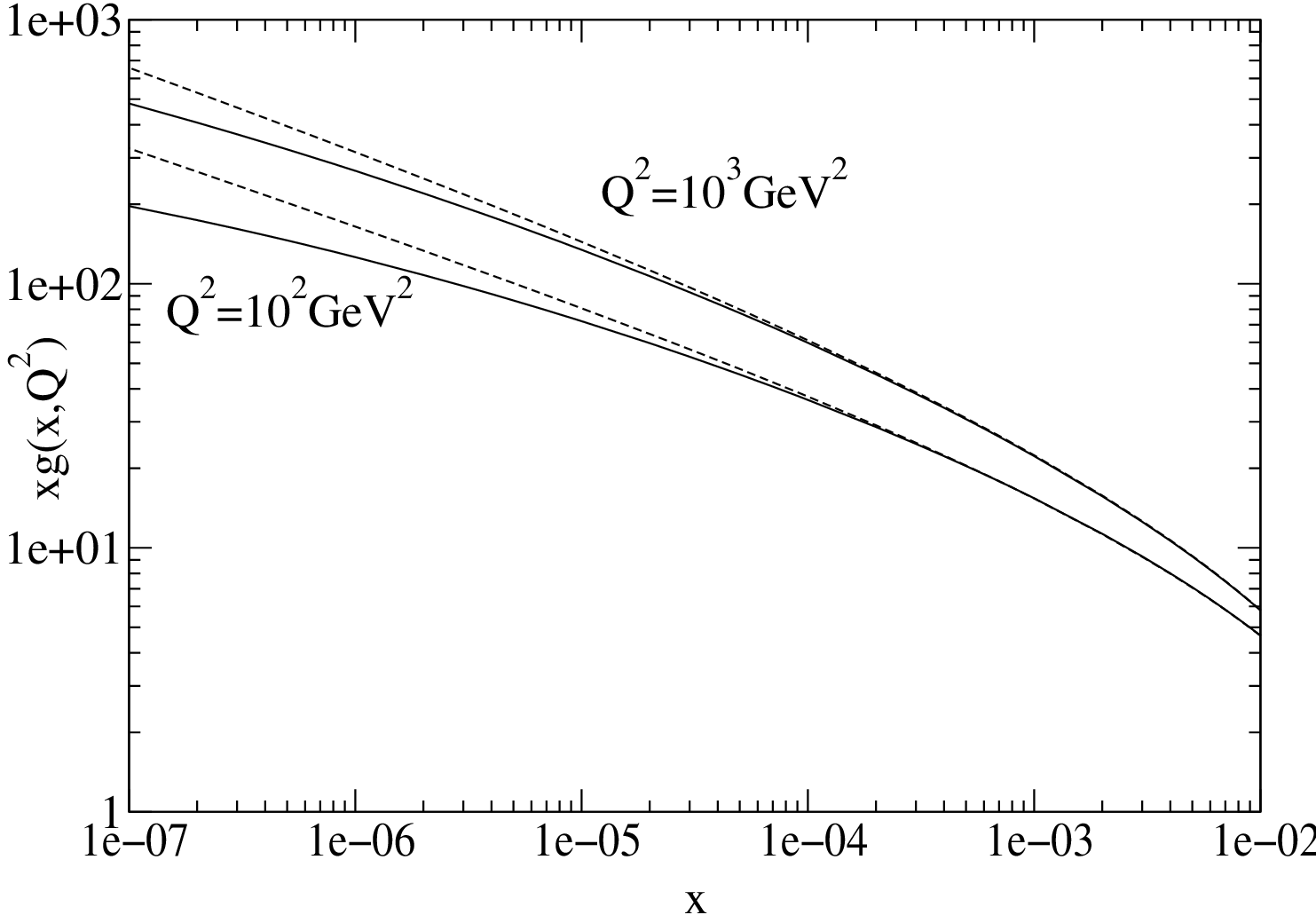}
\hfill\includegraphics[width=0.489\textwidth]{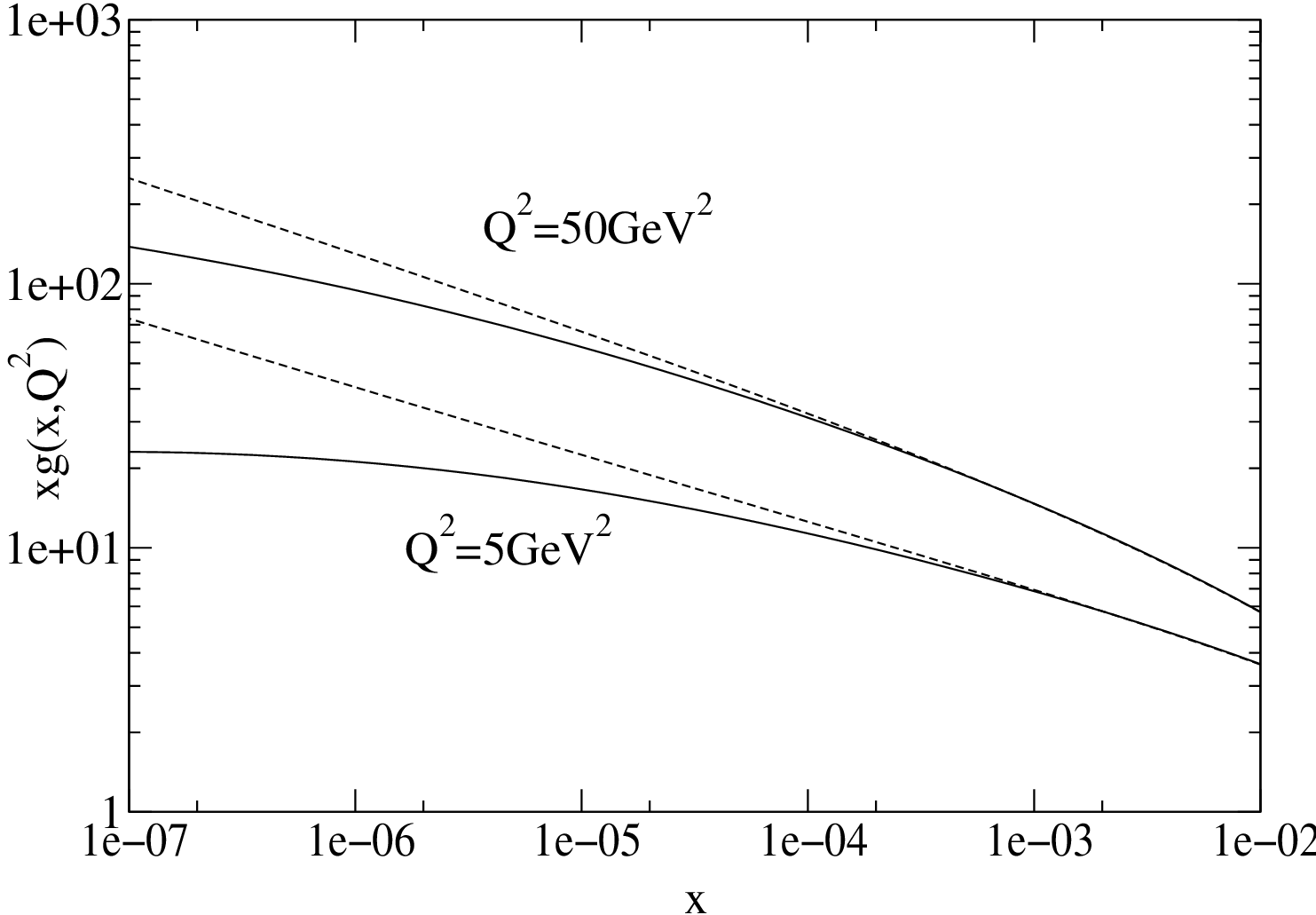}
\\} 
\caption{\em The integrated gluon distribution $xg(x,k^2)$ as a 
function of $x$ for  values of $Q^2 = 10^2 \gev^2$ and $Q^2 = 
10^3$ (left) and for $Q^2 = 5 \  \gev^2$ and $Q^2 = 
50$ (right) obtained from integrating $f(x,k^2)$Eq.(\ref{eq:kovev}). Dashed
lines   correspond to solution of linear BFKL/DGLAP evolution equation.} 
\label{fig:xg} 
\end{figure}

In Fig.~\ref{fig:xg} we show the integrated gluon   density given by Eq.~(\ref{eq:xg}).
The change from the power behaviour   at small $x$   is clearly
visible in the nonlinear case. Also the differences   between    the
distributions in the linear and nonlinear case seem to be more   pronounced 
for the quantity $xg(x,Q^2)$.   This is due to the fact that in   order to
obtain the gluon density $xg(x,Q^2)$   one needs to integrate over   
scales up to $Q^2$ including small values of $k^2$, where the     suppression
due to the nonlinear term is bigger.  
\subsection{The saturation scale $Q_s(x)$}  
 
\begin{figure}[t] 
\centering{   
\includegraphics[width=0.7\textwidth]{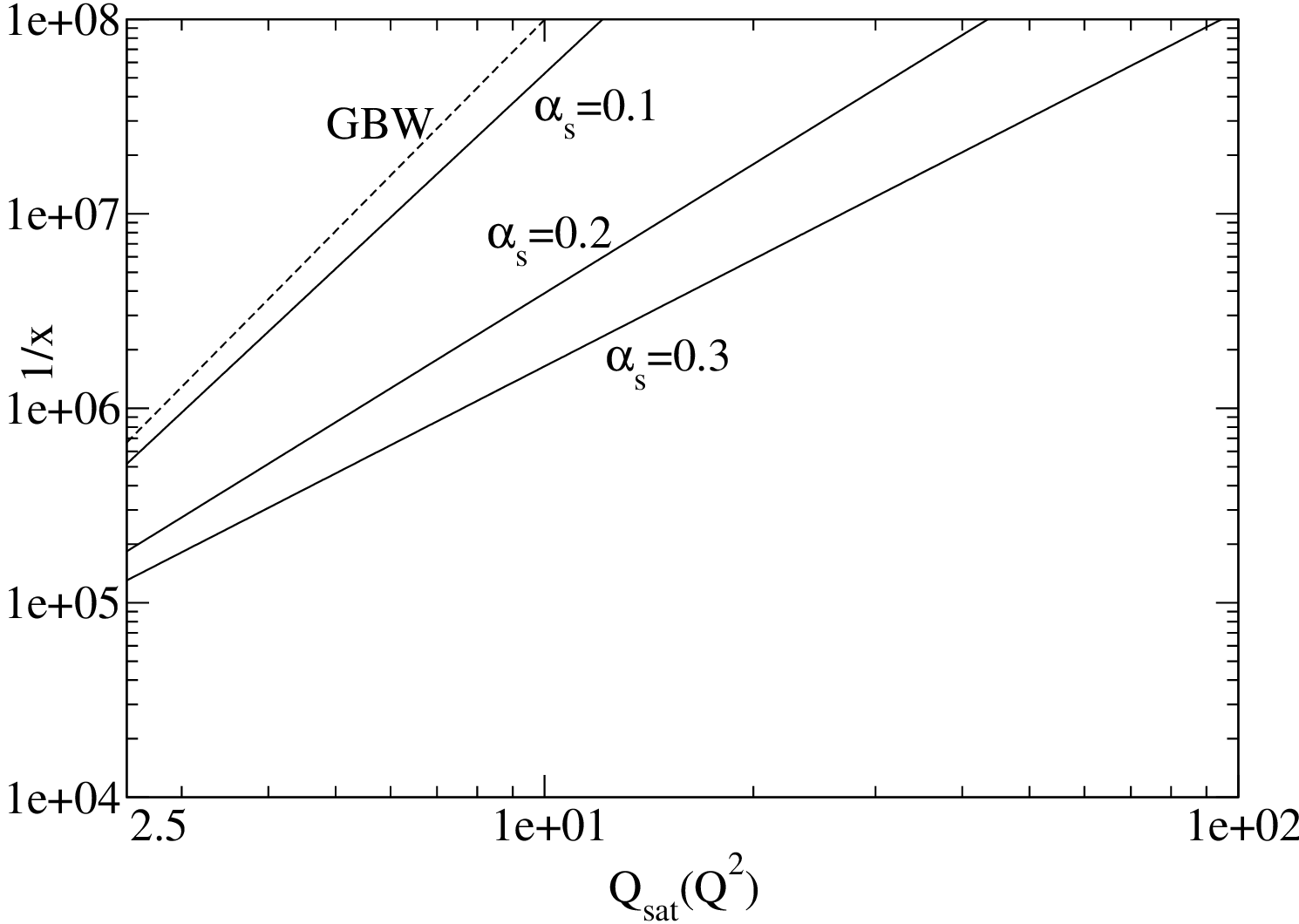} \\}   
\caption{\em Saturation
scale obtained from Eqs.~(\ref{eq:qsats},\ref{eq:omegas})  solid lines,
compared with saturation scale from GBW model \cite{GBW}.}  
\label{fig:qsat} 
\end{figure}

In order  to quantify the strength of the nonlinear term, one  introduces 
the  saturation scale $Q_s(x)$. It divides the space in $(x,k^2)$  
into regions of the dilute and dense partonic system. In the case when $k^2 <
Q_s^2(x)$ the solution    of the nonlinear BK equation exhibits the geometric scaling.
This  means that  it is dependent only     on  one variable $N(r,x)=N(r Q_s(x))$, or
in momentum space $\phi(k,x) = \phi(k/Q_s(x))$.   Recently, an analysis of the
saturation scale in the case of the model   with resummed NLL BFKL has been
performed \cite{MartinSat}.  There, the saturation scale was calculated
from the formula   \be   -\frac{d \omega(\gamma_c)}{d\gamma_c} \; = \;
\frac{\omega_s(\gamma_c)}{1-\gamma_c} \; ,  \label{eq:qsats} 
\ee 
which has been first derived in \cite{GLR} by the boundary condition of the wave front.
Formula (\ref{eq:qsats}) has been later rederived in \cite{MuTr,MuPe}.
The effective Pomeron intercept $\omega_s$   is a solution to the equation 
\be 
\omega_s(\gamma) \; = \; \bar{\alpha}_s \, \chi(\gamma,\omega_s) \; ,
\label{eq:omegas}
\ee 
where $\chi(\gamma,\omega)$ is  the kernel eigenvalue of the resummed model.
In our case the eigenvalue has the following form
\be
\chi(\gamma,\omega) \; = \; 2\Psi(1)-\Psi(\gamma)-\Psi(1-\gamma+\omega) +
\frac{\omega}{\gamma} \bar{P}_{gg}(\omega) \; .  \label{eq:chieigen}
\ee
The solution for the saturation scale  obtained from solving (\ref{eq:qsats},\ref{eq:omegas}) using eigenvalue (\ref{eq:chieigen})
is shown in Fig.~\ref{fig:qsat}
and gives $\lambda=\frac{\omega_s(\gamma_c)}{1-\gamma_c}=0.30,0.45,0.54$ for
three values of $\alpha_s=0.1,0.2,0.3$, respectively. These results are similar to
those obtained in \cite{MartinSat}. We compare our results with the saturation
scale   from the Golec-Biernat and W\"usthoff model. Normalisation of the
saturation scale is set to match GBW saturation scale at $x_0 = 0.41 \times
10^{-4}$.   

\begin{figure}[htb] 
\centering{ 
 \includegraphics[width=0.489\textwidth]{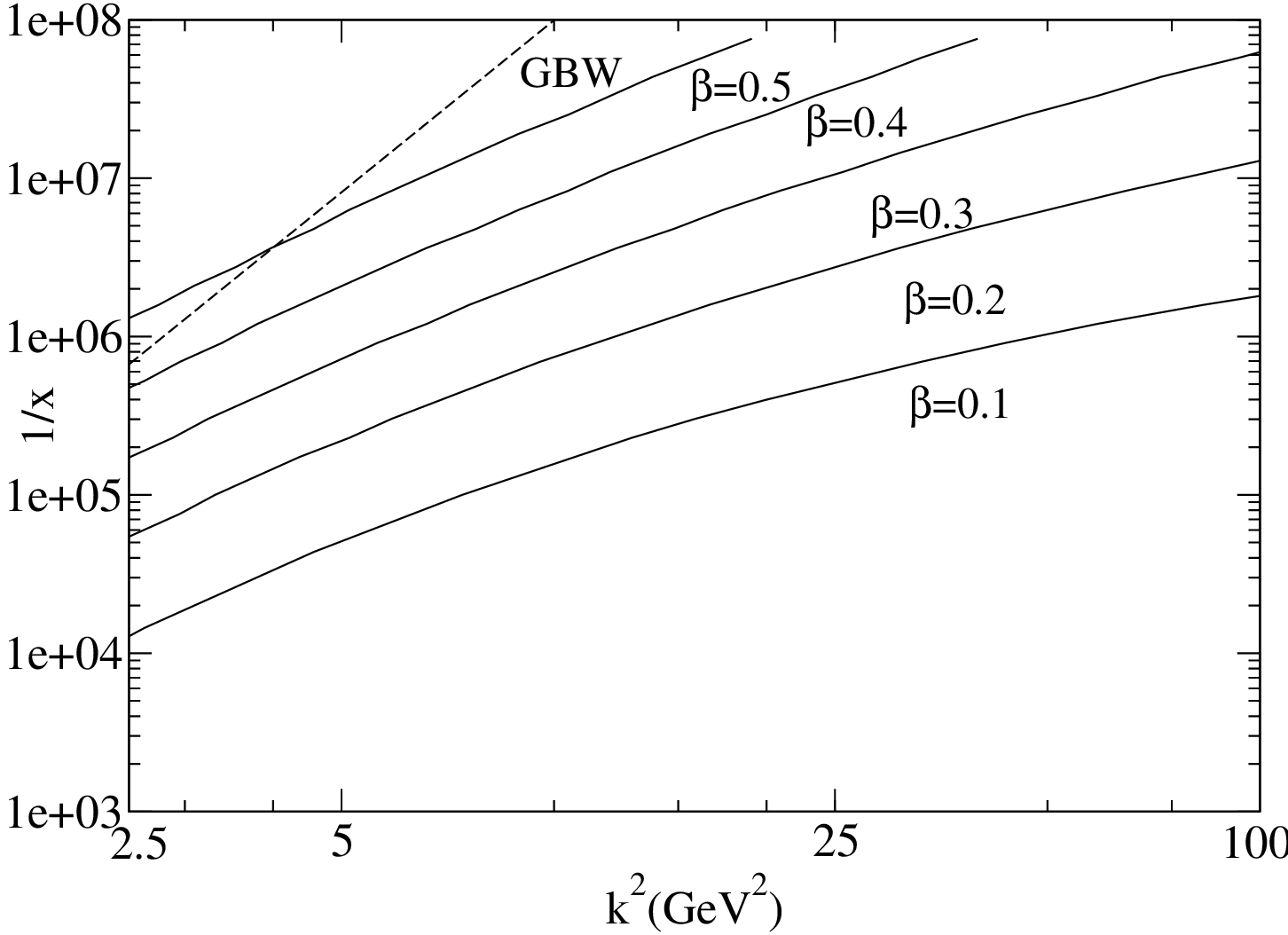}\hfill \includegraphics[width=0.489\textwidth]{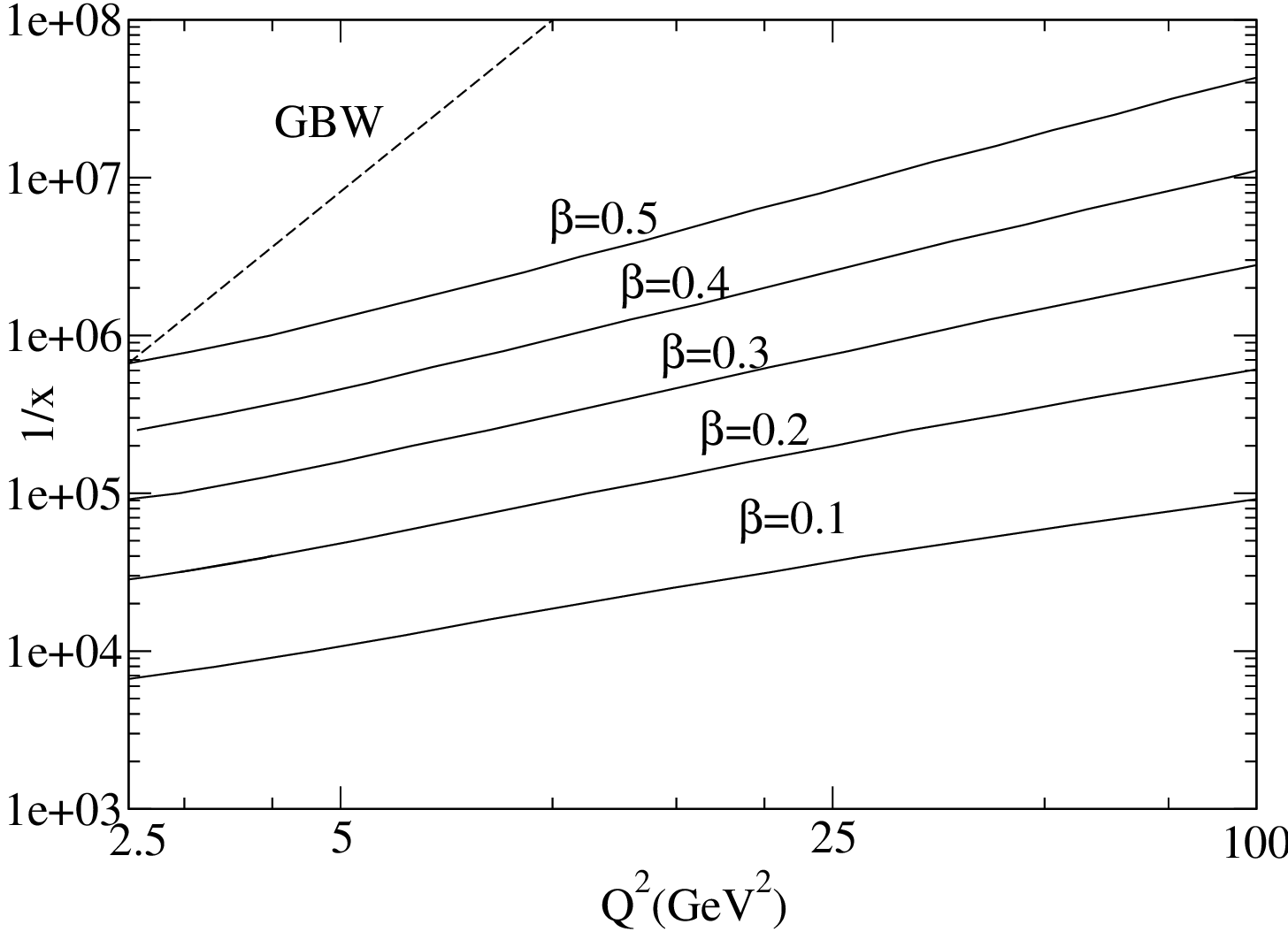} \\} 
\caption{\em Solid lines show contours of constant relative difference between solutions to linear and nonlinear equations, Eq.~(\ref{eq:qsat}).  
Lines from bottom to top correspond to $10\%,20\%,30\%,40\%,50\%$ difference. 
Left: Contours in the case of the unintegrated gluon distribution $f(x,k^2)$; 
right: contours in the case of the integrated gluon distribution $xg(x,Q^2)$. 
Dashed line in both case corresponds to the saturation scale from
Golec-Biernat and W\"usthoff model \cite{GBW}.}  
\label{fig:qsatbeta}  
\end{figure}

The saturation scale $Q_s(x)$ can be also obtained directly from the numerical solution to the nonlinear equation 
by locating, for example, the maximum of the momentum distribution of the unintegrated gluon density in the spirit
of method presented in Ref.~\cite{GBMS}.
For the purpose of phenomenology we  attempt here to  estimate the effect
 of the nonlinearity in a different, probably more quantitative way.
We  study the relative difference between the solutions to the linear and nonlinear equations 
\begin{equation}  
\frac{|f^{\rm lin}(x,\tilde{Q}_s(x,\beta)^2)-f^{\rm nonlin}(x,\tilde{Q}_s(x,\beta)^2)|}  
{f^{\rm lin}(x,\tilde{Q}_s(x,\beta)^2)} \; = \; \beta  
\label{eq:qsat}  
\end{equation}  
where $\beta$ is a constant of order $0.1-0.5$.  
Since this definition of the saturation scale is different from the one  
used in the literature and is likely to posses different $x$ dependence, we denote  it as $\tilde{Q}_s$.    
In Fig.~\ref{fig:qsatbeta}(left) we show a set $\tilde{Q}_s$ which are   solutions to Eq.~(\ref{eq:qsat}) for different choices of $\beta$ together with the  
saturation scale calculated from the original saturation model by Golec-Biernat and W\"usthoff \cite{GBW}. Solid lines given by Eq.(\ref{eq:qsat}) show where the nonlinear solution  
for the unintegrated gluon starts to deviate from the linear one by 
 $10\%,20\%,\dots,50\%$. It is interesting  that 
contours $\tilde{Q}_s(x)$ defined in  (\ref{eq:qsat}) have much stronger
$x$ dependence   than saturation scale  $Q_s(x)$ defined by
Eq.(\ref{eq:qsats}) and the one from  GBW model.   In particular 
$\tilde{Q}_s(x,\beta) \, > \, Q_s(x)$ for given $x$ (at very small values of $x$). 
This might be a hint that saturation corrections can  become  important much 
earlier (i.e. for lower energies) than it would be expected from the usual definition of the saturation scale $Q_s(x)$.   In Fig.~\ref{fig:qsatbeta}(right) 
we also show contours in the case of the integrated gluon distribution function, that is the solution to (\ref{eq:qsat}) with $f(x,k^2)$ replaced by $xg(x,Q^2)$. 
As already seen from the previous plot, Fig.~\ref{fig:xg}, the differences in the integrated gluon are more pronounced. For example in the case of $Q^2=25 \gev^2$ 
and $x\simeq 10^{-5}-10^{-6}$  we expect about $15\%$ to $30\%$ difference in the normalization.   Again, by looking solely at the position of the critical line, one would expect the nonlinear effects to be completely negligible in this region since at $x=10^{-6}$  the  corresponding $Q_s^2(x) \simeq 2.8 \gev^2$  (taking  $Q_s^2(x)=Q_{s,0}^2 (x/x_0)^{-\lambda}$ with  normalization  $Q_{s,0}^{2}=1 \gev^2$ at $x_0\simeq 4 \times 10^{-5}$ and $\lambda \simeq 0.28$, \cite{GBW}). This rough analysis shows that one cannot think of saturation scale 
as a definite  and sharp border between very dilute and dense system. 
 The transition 
between these two regimes appears to be rather smooth and the nonlinear term 
of the equation seems to have quite a large impact on the normalization 
even in the 'linear' regime defined as $Q^2 \gg Q_s^2(x)$.

In practice, the estimate of the saturation effects is even more complicated since the unintegrated gluon density has to be convoluted with some impact factor, and the integration over the range of scales must be performed.

\subsection{Dipole cross section $\sigma(r,x)$}  
  
It is interesting to see what is the behavior of the dipole cross section $\sigma(r,x)$ obtained from  
the unintegrated gluon density via Eq.~(\ref{eq:dipxsec}). In this calculation 
we assume that $\alpha_s$ is running with the scale $k^2$. 
 
Calculation of the dipole cross section requires the knowledge of the unintegrated gluon density for 
all scales $0<k^2<\infty$. 
Since in our formulation the unintegrated gluon density is known for 
$k^2>k_0^2$ we need to parametrise $f(x,k^2)$ for lowest values of $k^2<k_0^2$. We use the matching 
condition 
\be 
xg(x,k_0^2)=\int_{0}^{k{_0}^2}\frac{dk^2}{k^2}f(x,k^2)  \; ,
\ee 
and following \cite{GBBK} we assume that $f(x,k^2)\sim k^4$ for low $k^2$. 
This gives  (compare Eq.(\ref{eq:gluony}) )
\be 
f(x,k^2)=4N(1-x)^{\rho} k^4 \; .
\ee 


\begin{figure}[t] 
\centering{ 
 \includegraphics[width=0.7\textwidth]{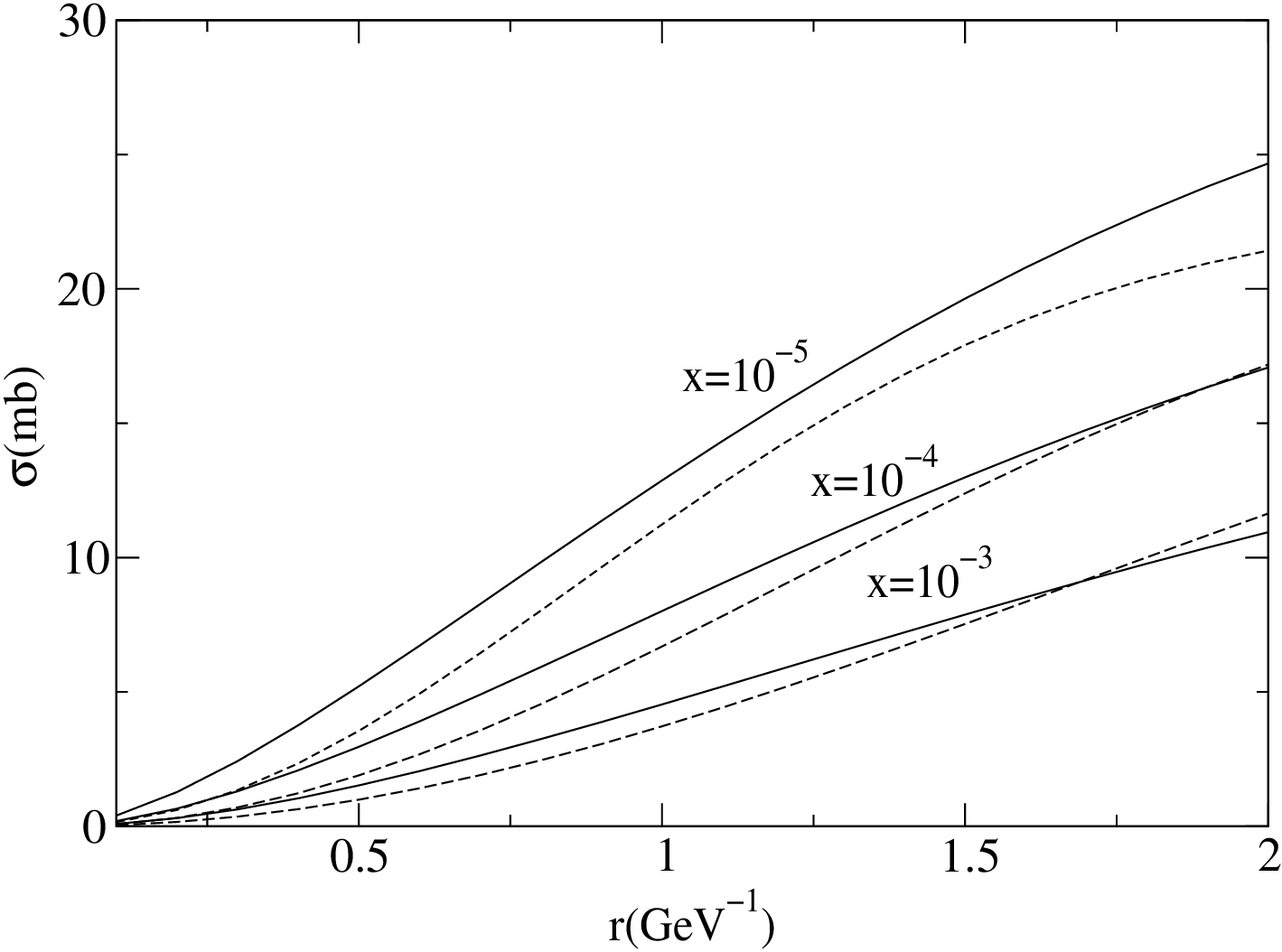}\\} 
\caption{\em The dipole cross section obtained from  
modified BK (solid line) compared to GBW dipole model (dashed line).} 
\label{fig:dipol} 
\end{figure}

In Fig.~\ref{fig:dipol} we present the dipole cross section as a function of the 
dipole size r for three values of $x=10^{-3},10^{-4}$ and $10^{-5}$. For comparison 
we also present  the dipole cross section obtained from GBW parametrisation. To 
be self-consistent, we cut the plot at $r=2 \ \gev^{-1}$ because we assumed in the 
derivation  of formula (\ref{eq:JKKK}) that the dipoles are small in comparison 
to the target size (we assume proton radius to be $4\, {\rm GeV}^{-1}$). 
This cut 
allows us   to obtain a model independent result since we observe that 
different  parametrisations of $f(x,k^2)$ for $k^2<k_0^2$ give essentially 
the  same  contribution for $r<2\, {\rm GeV}^{-1}$.  

We observe that our extraction of the dipole cross section gives similar result to the GBW parametrisation.
The small difference in the normalisation is probably due to the different values of $x_g$ which probe the gluon
distribution (or alternatively the dipole cross section).
In the GBW model the dipole cross section is taken at the value $x_g=x$ which is the standard Bjorken
 $x = Q^2/2 p\cdot q$. 
On the other hand, in the 
formalism presented in Ref.~\cite{KMS} 
one takes into account the exact kinematics (energy conservation) in the photon impact factor.
It is a part of the subleading effect in the impact factor and it increases the value of $x_g \sim 5 x$.
Therefore, in our formalism the normalization of the unintegrated gluon  is increased so that
the convolution with the impact factor and the resulting structure function remains the same.

\section{Conclusions}
In this paper we studied numerically the solutions to the  modified BK equation
in the approximation of the infinite and uniform target.
The modifications include the subleading corrections in $\ln 1/x$ which
are given by the kinematical constraint, DGLAP $P_{gg}$ splitting function and
the running  of the strong coupling. Since these corrections reduce significantly the value
of the BFKL intercept they also have large impact onto the behavior of the saturation scale 
and the normalization of the solution.  For example, we find that at $x=10^{-4}$ and $k^2=100 \, {\rm GeV}^2$
 the normalization of the unintegrated gluon distribution is reduced by about $30\%-40\%$ as compared with  the solution to the  unmodified BK equation. 
 
 We have studied the onset of the nonlinear corrections by observing the difference in the normalization of the linear
 and non-linear solutions. We observe that even though the solutions are in the nominally dilute regime, the
 normalization of the solution to the BK equation can be already strongly affected by the presence
 of the nonlinear term. This can have potential impact onto  the extrapolation of the parton distributions
 to lower values of $x$. We note however that, as long as the nonlinearities do not affect substantially the $k$ and $x$
 dependence of the solution, the linear equation can  probably be used with suitably  chosen boundary conditions.
 
 We have also computed the dipole cross section from this model and compared it with the GBW. We find that
 both models give similar results, with some small differences which can be atrributed to slightly different treatment
 of the photon impact factor.
 
 Finally, we stress that although the subleading corrections are included in this formalism  by using
 the available knowledge on the NLLx BFKL equation and the resummation procedures, they are taken into account only
 in the linear part of the BK equation. Consistent and complete treatment would require their inclusion in the 
 triple Pomeron vertex part,  which is so far known to LLx accuracy only.

\section*{Acknowledgments} 
We thank Jochen Bartels, Krzysztof Golec-Biernat, Hannes Jung, Misha Lublinsky, 
and Agustin Sabio-Vera for useful discussions.\\K.K. is 
supported by {\it Graduiertenkolleg}  Zuk\"unftige Entwicklungen in der
Teilchenphysik. This research is partially supported by the U.S. Department of Energy
Contract No. DE-AC02-98-CH10886 and by the Polish Committee for Scientific Research,
KBN Grant No. 1 P03B 028 28.


\end{document}